\documentclass[11pt]{article}

\usepackage[letterpaper,margin=1in]{geometry}
\usepackage[utf8]{inputenc} 
\usepackage[T1]{fontenc}    
\usepackage[colorlinks=true]{hyperref}       
\usepackage{url}            
\usepackage{booktabs}       
\usepackage{amsfonts}       
\usepackage{amsmath,dsfont,amsthm,mathtools,amssymb}
\usepackage{subcaption,enumitem}
\usepackage{thmtools,thm-restate}
\usepackage{tikz,soul,pgfplots}
\pgfplotsset{compat=1.18}
\usetikzlibrary{shapes,arrows,fit, calc}
\usetikzlibrary{shapes.misc, positioning,patterns}
\usetikzlibrary{decorations.pathreplacing}
\usetikzlibrary{arrows.meta, shapes,patterns.meta}

\DeclareMathOperator{\Var}{Var}
\DeclareMathOperator{\rank}{rank}
\DeclareMathOperator{\Bern}{Bernoulli}

\newcommand{\calR}{\mathcal{R}}

\newcommand{\calW}{\mathcal{W}}

\newcommand{\calZ}{\mathcal{Z}}

\newcommand{\bbE}{\mathbb{E}}
\newcommand{\bbF}{\mathbb{F}}
\newcommand{\bbR}{\mathbb{R}}

\newcommand{\indic}{\mathds{1}}

\newcommand{\eye}{\mathbf{I}}

\newcommand{\better}{\succeq}

\newcommand{\grapherr}[3]{\tilde{\mathsf{e}}_{#1}(#2,#3)}
\newcommand{\minerr}[2]{\mathsf{e}_{#1}(#2)}
\newcommand{\err}[1]{\mathsf{e}_{#1}}

\newtheorem{theorem}{Theorem}

\newtheorem{lemma}{Lemma}
\newtheorem{proposition}{Proposition}
\newtheorem{fact}{Fact}

\newtheorem{definition}{Definition}

\newcommand{\Df}[1]{\mathsf{D}_{#1}}
\newcommand{\Egamma}[1]{\mathsf{E}_{#1}}

\hypersetup{
	colorlinks,
	linkcolor=blue,
	citecolor=red,
	urlcolor=blue
}
\usepackage[
backend=biber,
style=alphabetic,
giveninits=true,   
maxbibnames=99,    
maxalphanames=4,   
minalphanames=3,   
date=year,        
doi=true,
url=true,
eprint=true,
isbn=false,
backref=true,
]{biblatex}
\expandafter\patchcmd\csname abx@ffd@*@eprint:arxiv\endcsname
{arXiv}{arXiv preprint}{}{}
\DeclareFieldAlias{eprint:arXiv}{eprint:arxiv}
\AtEveryBibitem{%
	\ifentrytype{inproceedings}{%
		\clearlist{publisher}%
		\clearlist{location}%
		\clearlist{organization}%
		\clearfield{venue}%
		\clearfield{series}%
	}{}%
	
	\iffieldundef{doi}
	{%
		\iffieldundef{eprint}{}{%
			\clearfield{url}%
			\clearfield{urlyear}%
		}%
	}
	{%
		\clearfield{eprint}%
		\clearfield{url}%
		\clearfield{urlyear}%
	}%
}
\DeclareFieldFormat[article,periodical]{volume}{%
	\bibstring{jourvol}\addnbspace#1}
\DeclareFieldFormat[article,periodical]{number}{%
	\bibstring{number}\addnbspace#1}
\renewbibmacro*{volume+number+eid}{%
	\printfield{volume}%
	\setunit*{\addcomma\space}%
	\printfield{number}%
	\setunit{\bibeidpunct}%
	\printfield{eid}}

\title{On the Optimality of Truncation Among Binary Linear Sketches} 

\author{%
	Robinson D.~H.~Cung,  Adway Girish, Emre Telatar \\
	{\small    School of Computer and Communication Sciences, EPFL} \\
	{\small   \texttt{\{robinson.cung,adway.girish,emre.telatar\}@epfl.ch} }
}

\begin{document}

	\pagenumbering{gobble}
	\maketitle

	\begin{abstract} 
		We study statistical inference and function evaluation problems in 
		which the input binary vector is available only through a binary linear 
		sketch, i.e., a fixed number of parity measurements.  We show that 
		under a certain monotonicity condition, simply truncating the input by 
		retaining a subset of its coordinates is optimal among all linear 
		sketches of the same size.  In particular, this applies to distributed 
		testing of correlations of binary vectors when they are of opposite 
		signs and evaluating unate Boolean functions on uniformly random inputs.
		Additionally, given that truncation is optimal, we immediately have 
		lower bounds for arbitrary binary linear sketches.  To illustrate this 
		application, we consider the one-way communication complexity of a 
		distributional graph connectivity problem on $n$ vertices and show that 
		truncation, and hence all binary linear sketches, requires $\Theta(n 
		\log n)$ bits to achieve constant error, while a nonlinear protocol 
		only requires $O(\log n)$ bits.
	\end{abstract}
	\tableofcontents
	
	\newpage
	\pagenumbering{arabic}

	\section{Introduction}
	\label{sec: intro}

	Linear sketches are a fundamental algorithmic tool to process 
	high-dimensional data, particularly when dealing with space and 
	communication constraints 
	\cite{alon1996moments,muthukrishnan2005streams,woodruff2014sketching}. A 
	linear sketch represents a vector $x$ of dimension $n$ by a vector $Ax$ (of 
	dimension $k \leq n$), the product of $x$ with a $k \times n$ matrix $A$.  
	Being linear, the (linear) sketch of $x+y$ can be easily obtained by 
	summing the sketches of $x$ and $y$.  This makes them particularly 
	attractive in communication problems as they allow sketches of distributed 
	observations to be easily 
	aggregated~\cite{cormode2005sketching,agarwal2013merge,ahn2012graph,ahn2012sketches},
	 and in streaming problems, where the sketch of a changing input can be 
	efficiently 
	updated~\cite{indyk2005moments,muthukrishnan2005streams,ganguly2008freq}.

	In this paper, we focus on linear sketches over $\bbF_2$ (which we call 
	\emph{binary linear sketches}; the words ``binary'' and ``linear'' may be 
	dropped when clear from context), as studied by \cite{kannan2018linear}.  
	Formally, a binary linear sketch of $x \in \bbF_2^n$ is the product $Ax$ 
	where the matrix $A$ has entries in $\bbF_2$ and matrix multiplication is 
	performed over $\bbF_2$.  Hence, a $k$-bit linear sketch consists of $k$ 
	parity measurements on the vector $x$. 
	The restriction to $\bbF_2$ is not as limiting as it might first appear. 
	Indeed, Kannan, Mossel, Sanyal,
	and Yaroslavtsev~\cite{kannan2018linear} explicitly write that they are 
	``unaware of any problem for which sketching over reals gives any advantage 
	over $\bbF_2$'', and note that all (then) known dynamic graph streaming 
	algorithms (except~\cite{kapralov2017dynamic}) can be seen as binary linear 
	sketches without losing space-optimality.  Despite their simple structure, 
	binary linear sketches capture the power of general streaming algorithms in 
	several settings~\cite{hosseini2019optimality}, in addition to being 
	studied in their own right for hashing and randomness 
	extraction~\cite{carter1977universal,alon1999hash,dhar2024hash}.
	
	We consider using such sketches for downstream statistical inference and 
	function evaluation tasks.  Among all $k$-bit linear sketches, the simplest 
	choice is to retain $k$ of the $n$ input coordinates, which we call 
	\emph{truncation}.  We ask the following general question:
	\begin{quote}\vspace*{-2pt}
		\emph{When is a simple truncation of the input the best binary linear 
		sketch?}\vspace*{-2pt}
	\end{quote}
	By input we refer to the observations that the statistical inference is 
	based on, or the evaluation point of the function, as the case may be.
	We give a monotonicity condition that guarantees the optimality of 
	truncation, and provide applications to distributed correlation testing and 
	Boolean function evaluation.  Lower bounds on truncation then also hold for 
	all linear sketches; we use this to establish the strict suboptimality of 
	linear sketches against nonlinear encoders in a distributional graph 
	connectivity problem with one-way communication.
	
	Linear maps are classical tools in coding 
	theory~\cite{Richardson_Urbanke_2008,guruswami2012essential} and compressed 
	sensing~\cite{donoho2006compressed}, where the goal is to recover the 
	entire input.  They are also useful to recover a function of the input:\\ 
	K\"orner and Marton \cite{korner_marton} showed that applying the same 
	binary linear sketch to two symmetrically correlated binary vectors is 
	optimal for recovering their modulo-2 sum; see \cite{tsujino2026structured} 
	for recent developments.  
	Since the modulo-2 sum is a sufficient statistic for a hypothesis test on 
	the \emph{correlation}, one might expect linear sketches to also perform 
	well for statistical inference problems involving correlation.  Indeed, 
	linear  sketches are known to improve upon classical random quantization 
	for distributed hypothesis testing between certain pairs of positive 
	correlations~\cite{haim_kochman_itw,haim_kochman_arxiv}.
	
	In our previous work~\cite{girish2026suboptimality}, however,
	we showed that
	when the correlations have equal magnitude
	and opposite signs, or when either correlation is zero,
	no binary linear sketch can do better than trivially retaining the same 
	subset of coordinates (and the latter is easily shown to be worse than 
	classical quantization, hence linear sketches are strictly suboptimal for 
	this problem).  
	We conjectured that truncation remains optimal among binary
	linear sketches for \emph{every} pair of correlations of opposite signs 
	(and possibly different magnitudes).
	In the present paper, we resolve this conjecture in the affirmative.

	This optimality of simple coordinate projections is reminiscent of the 
	Courtade--Kumar conjecture~\cite{kumar_courtade}, with a long line of works 
	(e.g.,~\cite{samorodnitsky2015,dictator_mi,eldan2023boolean}) over the last 
	decade culminating in three independent proofs earlier this 
	year~\cite{chen2026proofinformativebooleanfunction,ky2026dictatorsinformative,mahdavifar2026informativebitbeyondproof}.
	  It states that dictator functions are the ``most informative'', i.e., 
	among all functions mapping a uniform vector $X$ to a single bit,  
	truncation (to a single coordinate) has the highest mutual information with 
	a version of $X$ that has each coordinate flipped independently with a 
	fixed probability.   We study an analogous
	question for a more general decision-theoretic notion of optimality 
	(Blackwell dominance~\cite{blackwell_order}) and arbitrary sketch
	dimension $k$, but restricted to the class of binary linear sketches.
	
	A more closely related analogue is lossy compression
	using binary linear sketches. Ancheta~\cite{ancheta1978linear} showed that 
	among such sketches, truncation is optimal (achieving minimum asymptotic 
	expected reconstruction Hamming distance) for uniformly random binary 
	vectors (see also~\cite{massey1978joint}).  They conjectured that for 
	biased binary vectors, the optimal linear sketch is truncation followed by 
	lossless compression of the retained coordinates; this was also resolved 
	positively earlier this year~\cite{wu2026linear}.
	
	Before describing our results, it is worth clarifying the following:
	Much of the sketching literature chooses the matrix $A$ at random.  We 
	consider deterministic matrices as in, 
	e.g.,~\cite{nelson2014determ,li2018determ,kannan2018linear}.  Choosing a 
	random matrix can improve worst-case guarantees, but for a given input 
	distribution and an ``average-error'' performance metric, some fixed matrix 
	performs
	at least as well as any randomized choice~\cite{Kushilevitz_Nisan_1996}.  
	Our results hold for every fixed matrix and extend to settings where the 
	sketches are chosen via shared randomness that is independent of the input.
	
	\subsection{Results} \label{sec: results}
	
	We show that truncation is the optimal binary linear sketch for the 
	following two problems: distributed testing of correlations between binary 
	vectors observed at different terminals using sketches, and evaluating 
	unate Boolean functions from sketches (resp.~(i) and (ii) in the list 
	below).
	Beyond identifying the conceptually simplest scheme to be optimal, this 
	provides a method to establish the limitations of binary linear sketches.  
	In general, linear sketches can involve parities of large, overlapping 
	sets, whose analysis can be non-trivial and require specialized machinery 
	(e.g., using approximate Fourier dimensions~\cite{kannan2018linear}).  
	We illustrate the ease with which truncation yields lower bounds on all 
	such linear sketches using a toy graph connectivity testing problem ((iii) 
	below).  More precisely, we show the following. 
	
	\begin{itemize}[leftmargin=1.7em,topsep=2pt,itemsep=2pt] 
		\item[(i)] {\textbf{Distributed correlation testing.}}  In 
		Section~\ref{sec: DHT}, we consider the distributed correlation test 
		setting of \cite{haim_kochman_itw,girish2026suboptimality}.   Alice and 
		Bob observe the respective components of $n$ independent binary pairs 
		with uniform marginals and correlation $\rho_h$ under hypothesis $h \in 
		\{0,1\}$.  They send $k_A$ and $k_B$ binary linear measurements 
		respectively to a central tester.  When the correlations are of 
		opposite signs, i.e., $\rho_0 \rho_1 \leq 0$, we show 
		(Theorem~\ref{thm:linear}) that simply sending the same $k = 
		\min\{k_A,k_B\}$ coordinates at both agents achieves the optimal 
		type-I--type-II error trade-off among all binary linear encoders.  This 
		proves the opposite-correlation conjecture 
		from~\cite{girish2026suboptimality}.

		\item[(ii)] {\textbf{Evaluating unate functions from linear 
		sketches.}}   A Boolean function is unate if it can be made 
		coordinate-wise non-decreasing by flipping a fixed subset of its input 
		coordinates.  
		For a uniformly random binary input vector, we show in 
		Theorem~\ref{thm: unate} that truncation (to an appropriate set of $k$ 
		coordinates) is the best choice for evaluating the function at the 
		vector from $k$ binary linear measurements of it. That is, for any 
		rank-$k$ matrix $A$, the best estimate of $f(z)$ obtained from $Az$ is 
		no better than the estimate obtained from the coordinates of $z$ 
		indexed by $k$ linearly independent columns of $A$.  See 
		Section~\ref{sec: unate} for more details.

		\item[(iii)] {\textbf{Strict suboptimality of all linear sketches for 
		graph connectivity.}}  We consider the problem of one-way communication 
		for graph connectivity~\cite{sun2015tight} (see Section~\ref{sec: 
		graph} for background).  Fix a graph. Alice observes a uniformly random 
		subgraph $G$ of this graph and sends a message to Bob, who observes a 
		fixed subgraph $H$ that is unknown to Alice.  Bob must decide whether 
		$G \cup H$ is connected with at most small constant error probability 
		for every $H$.  We take the original graph to be a complete bipartite 
		graph with with $v$ left and $2^v$ right vertices (and hence $n = v + 
		2^v$ total vertices) and show that arbitrary encoders can achieve this 
		using $O(\log n)$ bits, while truncation (and hence all binary linear 
		sketches) requires $\Theta(n \log n)$ bits (Theorem~\ref{thm: graph}). 
	\end{itemize}
	
	\subsection{Technical overview} \label{ref: tech_overview}
	
	Our optimality proofs follow from the classical theory on comparing 
	statistical experiments due to 
	Blackwell~\cite{blackwell_order,blackwell1953equivalent}.  To show that 
	truncation is
	optimal for a hypothesis test, we show that any binary linear sketch can be 
	statistically simulated (as in \eqref{eqn: kernel})
	from the truncated input through a transition kernel that does not
	depend on the hypothesis.  Thus, any performance achieved by the sketch
	can also be achieved by truncation.  The existence of such a kernel is 
	characterized by data processing inequalities holding for all 
	$f$-divergences (see Section~\ref{sec: prelims}, Fact~\ref{fact: blackwell} 
	for a formal statement).
	
	Without loss of generality (w.l.o.g.), we can write any linear sketch of 
	$x$ with a rank-$k$ matrix $A$ in the form $x_I \oplus \tilde A x_{I^c}$, 
	for some $I \subseteq [n] \coloneqq \{1,\dots,n\}$ and matrix $\tilde A$ of 
	appropriate dimension, where $x_I$ denotes the subvector of $x$ indexed by 
	the elements of $I$ (by an invertible transformation applied to $Ax$).  We 
	compare $x_I$ against $x_I \oplus \tilde A x_{I^c}$.  In our most general 
	technical result, we show that under an ``opposite monotonicity'' 
	assumption on the joint distributions of the pair $(U,V)$ under the two 
	hypotheses, $U$ on its own is better (via the $f$-divergence 
	characterization) than the mixed statistic $U \oplus V$. The intuition is 
	that under the opposite monotonicity assumption, mixing $V$ with $U$ 
	weakens the evidence in favour of either hypothesis already present in 
	$U$.  
	
	We first study the correlation testing problem (Section~\ref{sec: DHT}), 
	where we consider sketches of the vector $X \oplus Y$, where $X$ and $Y$ 
	are the correlated observations of Alice and Bob respectively with 
	uniformly random marginals.  It is easy to check that the opposite 
	monotonicity assumption holds when the correlations are of opposite signs.  
	However, we give a direct proof that exploits the coordinates of the vector 
	being independent and identically distributed (i.i.d.), which reduces the 
	problem to checking that a single bit of appropriate correlations can be 
	simulated.  This also gives an explicit kernel to show Blackwell dominance, 
	as in the special cases considered in~\cite{girish2026suboptimality}.  In 
	fact, we prove the stronger statement that truncation cannot be improved by 
	mixing any (possibly nonlinear) function of any quantity that is 
	independent of the truncated coordinates (even if they carry useful 
	information about the hypothesis).

	As for the Boolean function evaluation setting (Section~\ref{sec: unate}), 
	we take the function value to be the ``hypothesis'' and unateness gives the 
	desired opposite monotonicity.  Finally, it is easy to check the graph 
	connectivity predicate (Section~\ref{sec: graph}) is indeed a unate 
	function, and hence, truncation is the best linear sketch.  We show that 
	truncation (and hence any linear sketch) requires communication of at least 
	a constant fraction of the number of edges, i.e., $\Theta(n \log n)$ bits 
	to achieve small error, as a small fraction of edges can only account for a 
	small fraction of the variance. 
	On the other hand, we give a simple nonlinear scheme using the fact that 
	most right vertices are non-isolated and that all remaining vertices form 
	one connected component, which uses only $O(\log n)$ bits to achieve small 
	constant error probability.
	
	\subsection{Open problems and future directions} \label{sec: future}
	
	The most cleanly stated open problem is to identify the exact set 
	$\calR_{k,n} \subseteq [-1,1]^2$ of correlation pairs $(\rho_0,\rho_1)$ for 
	which truncation is the optimal linear sketch for finite $k,n$.   We show 
	here that $\rho_0\rho_1 \leq 0$ is a sufficient condition for all $k \leq 
	n$, but \cite[Section 5]{girish2026suboptimality} shows through numerical 
	experiments that $\calR_{k,n}$ appears to shrink to the set of 
	opposite-sign correlations as $k,n \to \infty$.  Outside this region, does 
	there exist such an optimal binary linear sketch, and can we identify it?
	
	Further, what structural features of an inference problem allow for 
	truncation to be the optimal sketch?  We provide a sufficient monotonicity 
	condition, but it would be interesting to characterize a more general 
	property, e.g., through the Fourier structure of the conditional 
	distributions.   Such a characterization might prove useful in identifying 
	other settings where truncation remains optimal.
	
	More generally, can we identify how much communication is required to 
	compensate for restrictions (such as linearity) on the encoder?  For 
	evaluating $f(X \oplus Y)$ at uniformly random $X$ and 
	$Y$,~\cite{kannan2018linear} show that binary linear sketches can match the 
	performance of arbitrary one-way encoders if they are allowed a 
	constant-factor overhead in communication.  Our comparison pits truncation 
	against sketches of the exact same size without allowing for such 
	overheads.  It would be interesting to see if similar Blackwell comparisons 
	can be obtained for other ``simple'' sketches that can match arbitrary 
	encoders.
	Our graph connectivity example shows that the overhead factor can be 
	$\Omega(n)$, so a constant factor does not suffice in general, but can we 
	identify additional assumptions so that it does?  This would help clarify 
	when nonlinear encoders are essential, justifying their potentially greater 
	implementation cost.
	
	\section{Preliminaries} \label{sec: prelims}
	
	We follow standard notation; a full summary is in Appendix~\ref{app: 
	notation} for completeness. 
	
	\begin{definition}[Binary linear sketch, information set, truncation]
		A \emph{binary linear sketch} is a map $x\mapsto Ax$, where
		$A$ is a matrix with $n$ columns and entries in $\bbF_2$. 
		We call $I \subseteq [n]$ an
		\emph{information set} for a full-row-rank matrix $A \in \bbF_2^{k 
		\times n}$ if $|I|=k$ and $A_I$, the submatrix of $A$ with columns 
		indexed by $I$, is invertible. 
		We call the coordinate projection $x\mapsto x_I$
		\emph{truncation} to $I$.    
	\end{definition}
	
	No information is lost on discarding linearly dependent rows from the 
	sketch, so we assume henceforth w.l.o.g. that $A$ has full row-rank.
	For any information set $I$ of $A$, note that $A_I^{-1}Ax = x_I \oplus 
	\tilde A x_{I^c}$ with $\tilde A = A_I^{-1}A_{I^c}$, so the sketches $Ax$ 
	and $x_{I} \oplus \tilde A x_{I^c}$ are equivalent.  
	
	\begin{definition}[$f$-divergence]
		Given two probability distributions $P,Q$ on a finite alphabet $\calZ$ 
		and a convex function $f : (0, \infty) \to \bbR$ with $f(1) = 0$, we 
		define the \emph{$f$-divergence} between $P$ and $Q$ as
		\begin{equation*} \label{eqn: f-div}
			\Df{f}(P \,\|\, Q) \coloneqq \sum_{z} Q(z) 
			f\left(\frac{P(z)}{Q(z)}\right)
		\end{equation*}
		with the usual limiting conventions, i.e., we set $f(0) = \lim_{t 
		\downarrow 0} f(t)$, assign zero to terms with $P(z)=Q(z)=0$, and take 
		$Q(z)f(P(z)/Q(z))$ with $Q(z) = 0 < P(z)$  to mean $P(z) \lim_{t \to 
		\infty} f(t)/t$.
	\end{definition}
	An important property of 
	$f$-divergences~\cite{csiszar1967fdiv,Polyanskiy_Wu_2025} is that they are 
	jointly convex in their arguments.
	One particular choice of $f$ that will be of interest to us is the convex 
	function $f_{\gamma}(t) = [t-\gamma]_+ - [1-\gamma]_+$ for $\gamma \geq 0$, 
	where $[t]_+ \coloneqq \max\{t,0\}$, which yields the 
	\emph{$\Egamma{\gamma}$-divergence} (also called the ``hockey-stick 
	divergence''~\cite{sason2016fdiv,liu2017egamma}).  
	This forms an important family as it generates all $f$-divergences, i.e., 
	any $f$-divergence can be written as a non-negative linear combination 
	(possibly an integral) of $\Egamma{\gamma}$-divergences.
	
	\begin{definition}[$\Egamma{\gamma}$-divergence]
		Given two probability distributions $P,Q$ on a finite alphabet $\calZ$, 
		we define the \emph{$\Egamma{\gamma}$-divergence} between $P$ and $Q$ as
		\begin{equation*}
			\Egamma{\gamma}(P \,\|\, Q) \coloneqq \sum_z [P(z) - \gamma Q(z)]_+ 
			- [1-\gamma]_+.
		\end{equation*}
	\end{definition}
	
	Finally, our main technical tool is the notion of Blackwell dominance for 
	comparing statistics: we take $U$ to be better than $V$ for testing between 
	possible choices of the hypothesis $h$ if we can simulate $V$ from $U$ 
	without knowing $h$.  This has a clear interpretation: someone who observes 
	$U$ can also simulate $V$ locally and make any decision that someone who 
	observes only $V$ makes, and hence, cannot do worse.  We also have a 
	convenient characterization~\cite{blackwell_order} of exactly when this 
	partial order relation is satisfied, namely when all $f$-divergences 
	satisfy a \emph{data processing inequality}.
	
	\begin{definition}[Blackwell dominance]\label{def:blackwell}
		Suppose that $U$ and $V$ have distributions $P_U^h$ and
		$P_V^h$, respectively, under hypothesis $h \in \{0,1\}$.
		We say that $U$ \emph{Blackwell-dominates} $V$, denoted by
		$U \better V$, if there exists a single transition
		kernel $K$ such that for both $h = 0$ and $1$, we have
		\begin{equation}
			P_V^h(v)=\sum_u P_U^h(u)K(v\mid u). \label{eqn: kernel}
		\end{equation}
	\end{definition}
	
	\begin{fact}[Divergence characterization of Blackwell 
	dominance~\cite{blackwell_order}] \label{fact: blackwell}
		Under the binary hypothesis above, Blackwell dominance $U \better V$ is 
		equivalent  to the divergence data processing inequality
		\begin{equation}
			\Df{f}(P^1_U \,\|\, P^0_U) \geq \Df{f}(P^1_V \,\|\, P^0_V) 
			\label{eqn: blackwell_f-div}
		\end{equation}
		for every convex $f$ with $f(1) = 0$, or equivalently, 
		$\Egamma{\gamma}(P^1_U \,\|\, P^0_U) \geq \Egamma{\gamma}(P^1_V \,\|\, 
		P^0_V)$ for every $\gamma \geq 0$.
	\end{fact}

	\section{Distributed correlation testing} \label{sec: DHT}
	
	We study a communication game: Alice and Bob hold $X$ and $Y = X \oplus Z$ 
	respectively, where $X$ is uniform on $\bbF_2^n$ and $Z$ has i.i.d.\ 
	$\Bern(p_h)$ entries independent of $X$, under an unknown hypothesis $h \in 
	\{0,1\}$, with distinct $p_0, p_1 \in [0,1]$.  Equivalently, the pair 
	$(X,Y)$ is generated from a joint distribution $P_{XY}^{h}$ under which the 
	marginals are uniform but have correlation $\rho_h = 1-2p_h$ so that $X 
	\oplus Y$ has i.i.d.\ $\Bern(p_h)$ entries. Alice (resp.~Bob) sends a 
	$k_A$-bit (resp.~$k_B$-bit) message to a referee, who must determine $h$ 
	with a guess $\hat{H}$ (note that this is a random variable as it depends 
	on the observations).  Note that taking $k_B = n$ also includes the 
	``one-way'' communication model where Bob himself is the referee.  This 
	setting with uniform marginals correlated according to the hypothesis is 
	also a common 
	choice~\cite{hadar2019stoc,hadar2019isit,kochman_wang_itw,sahasranand2021correlation}
	 as it clearly highlights the distributed aspect of the problem: Alice and 
	Bob themselves have no information about the hypothesis, but with 
	communication, they are able to make a non-trivial decision.  
	
	This is an instance of the widely studied problem of distributed hypothesis 
	testing in information theory, where the primary focus has been on 
	characterizing the optimal trade-off between the error probability and the 
	amount of communication (see the classical survey~\cite{han_amari_survey} 
	for more background).  The optimal encoding is often a typicality-based 
	random 
	construction~\cite{ahlswede_csiszar,han_multiterminal,rahman_wagner_2012,wen2026exponent}
	 that has a prohibitively high space and memory requirement. 
	We focus on the correlation testing setup as described above, and we 
	consider a conceptually simple protocol where Alice and Bob both send 
	binary linear sketches of their observations, i.e., Alice sends $AX$ using 
	a rank-$k_A$ matrix $A$ and Bob sends $BY$ using a rank-$k_B$ matrix $B$.  
	In addition to being easy to implement~\cite{practical_DHT}, they emerge as 
	natural protocols to analyze since, as described in Section~\ref{sec: 
	intro}, linear sketches are known to be optimal to reconstruct the modulo-2 
	sum~\cite{korner_marton}, which is a sufficient statistic to test 
	correlation of binary vectors.  Moreover, for this correlation testing 
	problem, they were shown to outperform typicality-based constructions for a 
	subset of $p_0, p_1 \leq 
	\frac12$~\cite{haim_kochman_itw,haim_kochman_arxiv}. We show that when $p_0 
	\geq \frac12 \geq p_1$ (or by symmetry $p_1 \geq \frac12 \geq p_0$) 
	truncation is the best linear sketch, that is, no linear sketch outperforms 
	simply sending a prefix of the input.
	This is similar in spirit to the work of Amari~\cite{amari_fisher} (see 
	also \cite{chen_watanabe_yu2026}), who showed, by computing the Fisher 
	information of $(f(X), g(Y))$, that truncation is asymptotically optimal 
	among a class of ``linear-threshold encodings'' (which are different from 
	the class of linear sketches that we consider) for estimating the 
	correlation when $X,Y$ are nearly independent. 
	
	Our main result is stated formally below in Theorem~\ref{thm:linear}.  We 
	show that among all possible choices of linear sketches of Alice and Bob 
	(which could, in principle, be different), the optimal choice is for both 
	to truncate the observation.  Recall that ``optimality'' is in terms of 
	Blackwell dominance (as described in Section~\ref{sec: prelims}) which has 
	the following implication here.  In a hypothesis testing setup, there are 
	two kinds of errors, type-I and type-II, with probabilities $\err{h} = 
	P^h_{XY}\{\hat H \neq h\}$ for $h = 0,1$ respectively.  Every choice of 
	encoder (mapping $X$ and $Y$ to messages) and decoder (mapping messages to 
	a decision $\hat H$) achieves a particular error probability pair 
	$(\err{0}, \err{1})$.  Hence, if we show that $(X_{[k]}, Y_{[k]})$ 
	Blackwell-dominates $(AX, BY)$, then every $(\err{0}, \err{1})$ that can be 
	attained by the linear sketch $(AX,BY)$ using some decoder can also be 
	attained by the truncated message $(X_{[k]}, Y_{[k]})$.
	In particular,  $(X_{[k]}, Y_{[k]})$ can then do at least as well in both 
	type-I and type-II error probability as any linear sketch. 
	
	\begin{theorem}\label{thm:linear}
		Suppose that $p_0,p_1$ lie on opposite sides of $1/2$, i.e., either 
		$p_0 \leq 1/2 \leq p_1$ or $p_1 \leq 1/2 \leq p_0$.  Let $A$ be a 
		rank-$k_A$ matrix and $B$ be a rank-$k_B$ matrix.  Then, letting $k = 
		\min\{k_A, k_B\}$, we have
		\begin{equation*}
			(X_{[k]}, Y_{[k]}) \better (AX,BY),
		\end{equation*}
		for testing $X \oplus Y \sim \Bern(p_0)^{\otimes n}$ versus 
		$\Bern(p_1)^{\otimes n}$ with $X,Y$ uniform on $\bbF_2^n$.
	\end{theorem}
	
	This is exactly the statement of Conjecture 1 
	in~\cite{girish2026suboptimality}, where it was proved for the special 
	cases (1) $p_0 + p_1 = 1$ and (2) $p_0 = 1/2$ or $p_1 = 1/2$.  We prove 
	Theorem~\ref{thm:linear} in two steps:  we first show in 
	Proposition~\ref{prop:same} that both Alice and Bob using the same linear 
	sketch is always better (for any $p_0,p_1$) than using different sketches.  
	This is already known~\cite[Theorem 1]{girish2026suboptimality}, but we 
	repeat the proof as it is short and illustrates the idea of constructing an 
	explicit kernel to show Blackwell dominance.  Then, we show that when the 
	correlations under the hypotheses are of opposite signs, truncation is the 
	best choice of this same linear sketch.

	\begin{proposition}[Same linear sketch is always better]\label{prop:same}
		For any $p_0,p_1$ and any $n$-column matrices $A,B$, we have
		\begin{equation*}
			A(X \oplus Y) = AX \oplus AY \better (AX, BY) \quad \text{ and } 
			\quad B(X \oplus Y) = BX \oplus BY \better (AX, BY),
		\end{equation*}
		for testing $X \oplus Y \sim \Bern(p_0)^{\otimes n}$ versus 
		$\Bern(p_1)^{\otimes n}$ with $X,Y$ uniform on $\bbF_2^n$.
	\end{proposition}
	\begin{proof}
		Let $Z \coloneqq X \oplus Y$.  We prove the statement by constructing a 
		kernel to simulate $(AX, BY)$ from $A(X \oplus Y) = AZ \eqqcolon U$ 
		under hypothesis $h = 0,1$.
		Generate $W$ uniformly on $\bbF_2^n$, i.e., i.i.d.\ 
		$\Bern(1/2)^{\otimes n}$, independent of $Z$. Then $(AW \oplus U, BW)$ 
		has the same distribution as $(AX, BY)$. We are done since the 
		described kernel does not use the hypothesis $h$; we only generate a 
		uniform vector $W$, independent of everything else. The other side 
		holds similarly.
	\end{proof}

	Note that Proposition~\ref{prop:same} not only implies that the same linear 
	sketch should be applied by both Alice and Bob, but also that the modulo-2 
	sum $AX \oplus AY$ of the sketches is sufficient for this problem, in lieu 
	of the full pair $(AX, AY)$. This is perhaps unsurprising, as the sum $X 
	\oplus Y$ is a sufficient statistic for the hypothesis test over $(X,Y)$.  
	Thus, our original problem, namely identifying the best linear sketches 
	$(AX,BY)$ for testing $X \oplus Y \sim \Bern(p_0)^{\otimes n}$ versus 
	$\Bern(p_1)^{\otimes n}$,
	reduces to simply identifying the best linear sketch $AZ$ for testing $Z 
	\sim \Bern(p_0)^{\otimes n}$ versus $\Bern(p_1)^{\otimes n}$.
	
	We can assume w.l.o.g.\ that the rank-$k$ matrix $A$ is equal to $[\eye 
	\mid \tilde A]$ for some $k \times (n-k)$ matrix $\tilde A$.   Truncation 
	is the best linear sketch if and only if $Z_{[k]} \better Z_{[k]} \oplus 
	\tilde A Z_{[k]^c}$ for all $\tilde A$.  We will instead show a stronger 
	result in Proposition~\ref{prop:truncation}, namely that $Z_{[k]} \better 
	Z_{[k]} \oplus V$ for any $V$ independent of $Z_{[k]}$ but with possibly 
	different distributions $P^0_V$ and $P^1_V$ under the respective 
	hypotheses, when $p_0,p_1$ are on opposite sides of $1/2$.

	\begin{proposition}[Independent addition is bad]\label{prop:truncation}
		Suppose $U \sim \Bern(p_h)^{\otimes k}$ under hypothesis $h \in 
		\{0,1\}$.   We have that $U \better U \oplus V$ for every random $V \in 
		\mathbb{F}_2^k$ independent of $U$, with distribution $P_{V}^h$ under 
		hypothesis $h$ if and only if $p_0, p_1$ are on opposite sides of 
		$1/2$, i.e., $p_0 \leq 1/2 \leq p_1$ or $p_1 \leq 1/2 \leq p_0$.  
		Further, the kernel showing Blackwell dominance can be explicitly 
		constructed.
	\end{proposition}
	
	It is worth remarking that $V$ can carry useful information about the 
	hypothesis, so it is not trivial that adding $V$ to $U$ should not help 
	(and indeed, we will see in the proof that $U \better U \oplus V$ can fail 
	when $p_0,p_1$ are on the same side of $1/2$).
	Theorem~\ref{thm:linear} clearly follows as an immediate corollary of 
	Propositions~\ref{prop:same} and~\ref{prop:truncation}.  All that remains 
	is to prove Proposition~\ref{prop:truncation}, which follows from two short 
	lemmas. First, note that by Fact~\ref{fact: blackwell}, 
	Proposition~\ref{prop:truncation} is equivalent to the claim that for all 
	distributions $P^0_V, P^1_V$ over $\mathbb{F}_2^k$ and all $f$-divergences, 
	we have $\Df{f}( P^1_U \,\|\, P^0_U) \geq \Df{f}(P^1_{U \oplus V}  \,\|\, 
	P^0_{U \oplus V})$ if and only if $p_0, p_1$ are on opposite sides of 
	$1/2$.  We show in Lemma~\ref{lemma:reduction} that it is sufficient to 
	check this inequality for $V$ that is deterministic under both hypotheses, 
	and that it is sufficient to take one of these deterministic values to be 
	$0$.   We then construct an explicit transition kernel for a single bit in 
	Lemma~\ref{lemma:onebit} that suffices when $V$ is deterministic, putting 
	the two together to complete the proof of Proposition~\ref{prop:truncation} 
	below.
	
	\begin{lemma}[Reduction to deterministic vectors]\label{lemma:reduction}
		Let $U$ be a random vector on $\bbF_2^k$ with any distribution $P^h_U$  
		under hypothesis $h \in \{0,1\}$.  For an arbitrary deterministic $v 
		\in \bbF_2^k$, let $V_v$ denote the random variable that is 
		deterministically equal to $v$ under $h=1$ and equal to $0$ under $h = 
		0$.  If $U \better U \oplus V_v$ for every deterministic $v$, then we 
		also have $U \better U \oplus V$ for any $V$ independent of $U$ with 
		distribution $P^h_V$ under hypothesis $h = 0,1$.
	\end{lemma}
	\begin{proof}
		The assumption $U \better U \oplus V_v$ is equivalent to 
		\begin{equation*}\label{eq:deterministic_vector}
			\Df{f}( P^1_{U}  \,\|\, P^0_{U})  \geq  \Df{f}( P^1_{U \oplus v}  
			\,\|\, P^0_{U})
		\end{equation*}
		for every $f$-divergence (i.e., for all convex $f$ with $f(1) = 0$).
		Now for arbitrary $P^h_V$, $h = 0,1$, let $V \sim P_{V}^h$ be 
		independent of $U$, and let $\Df{f}$ be any $f$-divergence.  By the 
		joint convexity of $\Df{f}$, we have
		\begin{align}
			\Df{f}( P^1_{U \oplus V}  \,\|\, P^0_{U \oplus V})   & \leq 
			\sum_{v_0,v_1}  P^0_V(v_0)P^1_V(v_1)\,  \Df{f}( P^1_{U \oplus v_1} 
			\,\|\, P^0_{U \oplus v_0} ) \nonumber \\
			&\leq \max_{v_0, v_1} \Df{f}( P^1_{U \oplus v_1} \,\|\, P^0_{U 
			\oplus v_0} ) = \max_{v} \Df{f}( P^1_{U \oplus v} \,\|\, P^0_{U} ), 
			\label{eqn: det-red_last}
		\end{align}
		with the equality obtained by applying the bijection $u \mapsto u 
		\oplus v_0$ to both arguments, which does not change $\Df{f}$.  The 
		term on the right-hand side of \eqref{eqn: det-red_last}, by 
		assumption, is at most $\Df{f}( P^1_{U}  \,\|\, P^0_{U})$, and hence, 
		applying Fact~\ref{fact: blackwell} again gives the desired conclusion.
	\end{proof}

	\begin{lemma}[Single-bit simulation]\label{lemma:onebit}
		Suppose that the single bit $U$ has distribution $\Bern(p_h)$ under 
		hypothesis $h \in \{0,1\}$.  Then, we have $U \better U \oplus h$ if 
		and only if $p_0,p_1$ are on opposite sides of $1/2$, i.e., $p_0 \leq 
		1/2 \leq p_1$ or $p_1 \leq 1/2 \leq p_0$.
	\end{lemma}
	\begin{proof}
		We want to construct a transition kernel $K$ that maps $\Bern(p_0)$ to 
		itself, but $\Bern(p_1)$ to $\Bern(1-p_1)$ as in \eqref{eqn: kernel}.  
		Writing this in matrix notation, we want to find a stochastic matrix 
		$K=\begin{bmatrix}a&1-a\\ b&1-b\end{bmatrix}$ with $a,b \in [0,1]$ so 
		that
		\begin{align*}
			\begin{bmatrix}1-p_0&p_0\\ 
			1-p_1&p_1\end{bmatrix}\begin{bmatrix}a&1-a\\ 
			b&1-b\end{bmatrix}=\begin{bmatrix}1-p_0&p_0\\ 
			p_1&1-p_1\end{bmatrix}.
		\end{align*}
		Since both sides have rows summing to $1$, it suffices to check 
		equality for the first column, i.e.,
		\begin{align*}
			\begin{bmatrix}1-p_0&p_0\\ 1-p_1&p_1\end{bmatrix}\begin{bmatrix}a\\ 
			b\end{bmatrix}=\begin{bmatrix}1-p_0\\ p_1\end{bmatrix}.
		\end{align*}
		As $p_0 \neq p_1$, this has the unique solution $a = 
		p_1(1-2p_0)/(p_1-p_0)$, $b = (1-p_0)(2p_1-1)/(p_1-p_0)$.  It is easy to 
		check that they both lie in $[0,1]$ if and only if $p_0,p_1$ are on 
		opposite sides of $1/2$ (including $1/2$ itself), as claimed.
	\end{proof}
	
	We now put these lemmas together to complete the proof of 
	Proposition~\ref{prop:truncation}.
	
	\begin{proof}[Proof of Proposition~\ref{prop:truncation}]
		If $p_0,p_1$ are strictly on the same side of $1/2$, then taking $V = 
		V_v$ with $v = (1,0,\dots,0)$, we have, by Lemma~\ref{lemma:onebit}, 
		that $U \better U \oplus V_v$ if and only if $p_0,p_1$ are on opposite 
		sides of $1/2$.
		
		As for sufficiency, suppose $p_0, p_1$ are on opposite sides of $1/2$. 
		By Lemma~\ref{lemma:reduction}, it is enough to show that $U \better U 
		\oplus V_v$ for every deterministic $v \in \bbF_2^k$ (recall that $V_v$ 
		is deterministically equal to $v$ under $h=1$ and equal to $0$ under $h 
		= 0$). As the coordinates of $U$ are independent, we can simulate each 
		of them separately, as follows.
		For the coordinates at which $v$ has entry $0$, there is nothing to 
		do.  For the coordinates at which $v$ has entry $1$, use the kernel $K$ 
		from Lemma~\ref{lemma:onebit}, and we are done.  Let $K_v$ denote this 
		channel simulating $U \oplus V_v$ from $U$ for a deterministic $v$.  
		
		We now show how this also gives an explicit kernel from $U$ to $U 
		\oplus V$ for arbitrary distributions $P^h_V$, $h = 0,1$.  Let $v_0$ 
		and $v_1$ be two samples drawn from distributions $P^0_V$ and $P^1_V$ 
		respectively, with any coupling between them.  Pass $U$ through kernel 
		$K_{v_0 \oplus v_1}$, and add $v_0$ to the resulting vector.  The 
		resulting vector has distribution $P^h_{U \oplus v_h}$ under hypothesis 
		$h$.  Averaging this over all sampled $v_0 \sim P^0_V$ and $v_1 \sim 
		P^1_V$ gives a vector with distribution $P^h_{U \oplus V}$ when the 
		hypothesis is $h$, and we are done.  For the special cases (1) $p_0 + 
		p_1 = 1$ and (2) $p_0 = 1/2$ or $p_1 = 1/2$ applied to $(U,V) = 
		(Z_{[k]}, \tilde AZ_{[k]^c})$ for $Z \sim \Bern(p_h)^{\otimes n}$, this 
		recovers the kernels in~\cite{girish2026suboptimality} for some 
		coupling of $v_0,v_1$ above.
	\end{proof}

	We have thus shown that truncation is the best linear sketch for the 
	distributed testing for correlations of opposite signs from i.i.d.\ 
	samples.  Importantly, we require the independence of $Z_{I}$ and $Z_{I^c}$ 
	for $I = [k]$ under each hypothesis to use 
	Proposition~\ref{prop:truncation}.  In the next section, we consider a 
	Boolean function evaluation problem, where we no longer have this 
	independence.  Nonetheless, we show that a certain ``opposite 
	monotonicity'' assumption on the joint distribution of $Z_I$ and $Z_{I^c}$ 
	suffices, generalizing Proposition~\ref{prop:truncation}.

	\section{Evaluating unate Boolean functions from sketches} \label{sec: 
	unate}
	
	Let $Z$ be uniform on $\bbF_2^n$ and suppose that we wish to evaluate 
	$f(Z)$ without observing $Z$ directly.  Instead, we are given a rank-$k$ 
	linear sketch $AZ$.  Previous 
	work~\cite{friedgut1998juntas,kannan2018linear,yaroslavtsev2019approx} has 
	largely focused on characterizing the trade-off between the size of the 
	sketch and the approximation error through assumptions on the Fourier 
	structure of the target function $f$.  Instead, we show that as long as $f$ 
	satisfies a monotonicity property, namely that it is \emph{unate} (as 
	defined below), no rank-$k$ linear sketch performs better than truncation 
	to some $k$ input coordinates (which, naturally, depends on $f$).
	
	This can be viewed as a binary hypothesis testing problem, with the two 
	hypotheses corresponding to $f(Z) = 0$ and $f(Z) = 1$.  The conditional 
	distributions of $Z$ under these two hypotheses have different supports (as 
	$f(z)$ is necessarily exactly one of the values 0 and 1 for all $z$), and 
	hence, we can achieve zero error with access to $Z$ (this is a rewording of 
	the trivial observation that we can exactly evaluate $f(Z)$ if we know 
	$Z$).  However, the conditional distributions of an observation $S$ based 
	on $Z$, such as a linear sketch $AZ$, induced by the two hypotheses may not 
	have disjoint supports, making the problem non-trivial.

	\begin{definition}[Unate function]
		A function $f : \bbF_2^n \to \{0,1\}$ is \emph{unate} if there exists a 
		fixed $d \in \bbF_2^n$ such that the map $z \mapsto f(z \oplus d)$ is 
		non-decreasing in each coordinate.
	\end{definition}
	Equivalently, at each coordinate, a unate function is either non-decreasing 
	or non-increasing when all other coordinates are held fixed. 
	This includes all monotone Boolean functions, as well as functions obtained 
	from them by negating a fixed subset of the input variables.  Unate 
	functions are well-studied in switching theory~\cite{mcnaughton_unate}; 
	common examples include conjunctions and disjunctions (combinations 
	obtained by AND and OR operations, but with each coordinate negated 
	everywhere or nowhere), weighted threshold functions, tribes, and monotone 
	graph functions such as connectivity (see the classical 
	reference~\cite{ODonnell_2014} for more examples).
	
	For an observation $S$ obtained from $Z$, let
	\begin{equation}
		\minerr{f}{S} \coloneqq \min_g \Pr\{g(S) \neq f(Z)\}  \label{eqn: 
		min_err}
	\end{equation}
	denote the minimum probability of error in evaluating $f(Z)$ from $S$.  The 
	result below shows that among all possible linear sketches $S$ of $Z$, 
	truncation minimizes this error probability.  Further, truncation is also 
	optimal for any decision problem in which the unknown quantity is $f(Z)$ 
	and must be estimated from a linear sketch of $Z$.  
	
	\begin{theorem}[Evaluating unate functions] Let $Z$ be uniform on 
	$\bbF_2^n$, and let $f$ be a unate Boolean function.  For any rank-$k$ 
	binary matrix $A$ with information set $I$ of size $k$, we have $Z_I 
	\better AZ$ for any downstream experiment based on the ``hypothesis'' 
	$f(Z)$.  In particular, an optimal binary linear sketch of a given rank $k$
		for evaluating $f(Z)$ can be chosen to be a coordinate projection, 
		i.e., with $\minerr{f}{S}$ as defined in \eqref{eqn: min_err} denoting 
		the smallest error possible in evaluating $f$ from observation $S$, we 
		have, for any $k \in [n]$,
		\begin{equation*}
			\min_{A\, :\, \rank(A) = k} \minerr{f}{AZ}  =  \min_{I\, :\, |I| = 
			k} \minerr{f}{Z_I}.
		\end{equation*} \label{thm: unate}
	\end{theorem}
	To prove the theorem, we use the following generalization of the truncation 
	result from the previous section.  Note that conditioning on the function 
	value destroys the independence of coordinates, which was necessary for 
	Proposition~\ref{prop:truncation}. Instead, Proposition~\ref{prop: 
	gen_trunc} only requires the joint distributions (conditioned on the 
	function value) to satisfy a certain coordinate-wise monotonicity property, 
	which, as we shall see, follows from the unateness of $f$.
	
	\begin{proposition}[Addition is bad under opposite monotonicity]
		Let $U,V \in \bbF_2^k$ have joint distribution $P^h_{UV}$ under 
		hypothesis $h = 0,1$.  Suppose that for each fixed $v$, $P^0_{UV}(u,v)$ 
		is non-increasing in each coordinate of $u$, and $P^1_{UV}(u,v)$ is 
		non-decreasing in each coordinate of $u$.  Then, $U \better U \oplus V$ 
		for testing $P^0$ versus $P^1$.   \label{prop: gen_trunc}
	\end{proposition}
	\begin{proof}
		By Fact~\ref{fact: blackwell}, it suffices to compare the 
		$\Egamma{\gamma}$-divergences for every $\gamma\geq 0$, i.e., we want 
		to show that $\Egamma{\gamma}(P^1_{U} \,\|\, P^0_{U}) \geq 
		\Egamma{\gamma}(P^1_{U \oplus V} \,\|\, P^0_{U\oplus V})$.  Let 
		$U_{-\ell} \in \bbF_2^{k-1}$ denote the vector $U$ with its $\ell$-th 
		coordinate removed (and similarly $V_{-\ell}$ is $V$ without 
		$V_\ell$).  We first show that for any arbitrary function $W_\ell$ 
		taking values in some alphabet $\calW$ of the remaining coordinates 
		$(U_{-\ell}, V_{-\ell})$, replacing $U_\ell$ with $U_\ell \oplus 
		V_\ell$ cannot increase the desired $\Egamma{\gamma}$-divergence.  That 
		is, we show that
		\begin{equation}
			\Egamma{\gamma}(P^1_{U_\ell, W_\ell} \,\|\, P^0_{U_\ell,W_\ell}) 
			\geq 
			\Egamma{\gamma}(P^1_{U_\ell \oplus V_\ell, W_\ell} \,\|\, 
			P^0_{U_\ell \oplus V_\ell, W_\ell}), \label{eqn: one-step-ordering}
		\end{equation}
		where $W_\ell$ is any function of $(U_{-\ell}, V_{-\ell})$.
		We apply \eqref{eqn: one-step-ordering} successively by taking $W_1 = 
		U_{-1}$, $W_k = U_{-k} \oplus V_{-k}$, and at step
		$\ell = 2,\dots,k-1$,
		\begin{equation*}
			W_\ell =
			(U_1 \oplus V_1,\dots,U_{\ell-1}\oplus V_{\ell-1},
			U_{\ell+1},\dots,U_k).
		\end{equation*}
		It is easy to check that with this choice of $W_\ell$, the right-hand 
		side at each step is the left-hand side at the next step, starting from 
		$U$ on the left at $\ell = 1$ and ending with $U \oplus V$ on the right 
		at $\ell = k$.
		It therefore remains to prove
		\eqref{eqn: one-step-ordering} for arbitrary choices of $W_\ell$.  
		
		To see this, fix $w \in \calW$ and define $a_{ij}$ for $i,j = 0,1$ as
		\begin{equation*}
			a_{ij} = P^1_{UV}(U_\ell = i, V_\ell = j, W_{\ell} = w) - \gamma 
			P^0_{UV}(U_\ell = i, V_\ell = j, W_{\ell} = w).
		\end{equation*}
		By the (coordinate-wise opposite) monotonicity assumption, we have that 
		$a_{1j} \geq a_{0j}$ for both $j = 0,1$, as marginalizing over 
		$(U_{-\ell}, V_{-\ell})$ that have $W_{\ell} = w$ preserves 
		monotonicity. 
		
		Now we evaluate the left-hand side of \eqref{eqn: one-step-ordering} in 
		terms of $a_{ij}$'s.  To obtain the joint distribution of $(U_\ell, 
		W_\ell)$, we have to marginalize $a_{ij}$ over $V_{\ell} = 0,1$ in each 
		term, and hence the contribution of the $W_\ell = w$ term to the 
		left-hand side is\vspace*{-2pt}
		\begin{equation}
			[a_{00} + a_{01}]_+ + [a_{10} + a_{11}]_+. \label{eqn: 
			lhs}\vspace*{-2pt}
		\end{equation}
		Similarly, for the right-hand side, we require the joint distribution 
		of $(U_\ell \oplus V_\ell, W_\ell)$, for which we marginalize over 
		$(U_\ell, V_\ell)$ with the same value of $U_\ell \oplus V_\ell$, i.e., 
		$U_\ell = V_\ell$ in one term and $U_\ell \neq V_\ell$ in the other.  
		Hence, the contribution of the $W_\ell = w$ term to the right-hand side 
		is\vspace*{-2pt}
		\begin{equation}
			[a_{00} + a_{11}]_+ + [a_{01} + a_{10}]_+. \label{eqn: 
			rhs}\vspace*{-2pt}
		\end{equation}
		
		Thus, we are done if we can show that the expression in \eqref{eqn: 
		lhs} is at least that in \eqref{eqn: rhs}.  But this follows from the 
		convexity of the map $z \mapsto [z]_+$: set $x = a_{00} + a_{01}$ and 
		$y = a_{10} + a_{11}$.  If $x = y$ there is nothing to do (as 
		monotonicity forces $a_{00} = a_{10}$ and $a_{01} = a_{11}$). If $x 
		\neq y$ then set $\lambda = (a_{10} - a_{00})/(y-x)$.  Note that by the 
		monotonicity assumption, $\lambda \in [0,1]$.  The left-hand side 
		\eqref{eqn: lhs} is then $[x]_+ + [y]_+$, while the right-hand side 
		\eqref{eqn: rhs} is 
		\begin{equation*}
			[\lambda x + (1-\lambda) y]_+ + [(1-\lambda)x + \lambda y]_+,
		\end{equation*}
		which is upper bounded by $[x]_+ + [y]_+$ by the convexity of $z 
		\mapsto [z]_+$ applied to each term.  As this holds for every $w$, we 
		have proved \eqref{eqn: one-step-ordering}, and this completes the 
		proof.
	\end{proof}
	
	Proposition~\ref{prop: gen_trunc} also immediately recovers 
	Theorem~\ref{thm:linear}.  Indeed, writing $A = [\eye \mid \tilde A]$ 
	w.l.o.g.\ (as the correlation testing problem is symmetric in input 
	coordinates), we can take $U = Z_{[k]}$ and $V = \tilde A Z_{[k]^c}$ in 
	Proposition~\ref{prop: gen_trunc}.  It is immediately clear that when the 
	Bernoulli biases are on opposite sides of $1/2$, the required monotonicity 
	assumption holds: $U$ and $V$ are independent under both hypotheses, so the 
	joint probability being increasing/decreasing in $Z_i$ depends on the bias 
	being greater/smaller than $1/2$ respectively.
	We thus have $U = Z_{[k]} \better U \oplus V = AZ$ as a direct consequence.
	
	\begin{proof}[Proof of Theorem~\ref{thm: unate}]
		Assume that $f$ is not constant, as the constant case is trivial. For 
		any matrix $A$ with information set $I$ of size $k$, take $U = Z_I$ and 
		$V = \tilde A Z_{I^c}$ with $\tilde A = A_I^{-1}A_{I^c}$.  It suffices 
		to check that the joint distributions $P^h$ of $(U,V)$ conditioned on 
		$f(Z) = h$ for $h = 0,1$ respectively satisfy the required 
		coordinate-wise monotonicity properties.  Then, as $Z$ is uniform on 
		$\bbF_2^n$, we have the joint distributions
		\begin{equation*}
			P^h_{UV}(u,v) = \frac{2^{-n}}{\Pr\{f(Z) = h\}} \left|\{w \in 
			\bbF_2^{n-k} : \tilde Aw = v, \  f(u,w) = h\}\right|, 
		\end{equation*}
		where $f(u,w)$ denotes the evaluation of $f$ at $z$ with $z_I = u$ and 
		$z_{I^c} = w$.
		Note that we can write the above cardinality for $h = 0,1$ respectively 
		as $ \sum_{\substack{w \in \bbF_2^{n-k} \,:\, \tilde Aw = v}}  (1 - 
		f(u,w))$ and $ \sum_{\substack{w \in \bbF_2^{n-k} \,:\, \tilde Aw = 
		v}}  f(u,w)$.  Hence, for each coordinate of $u$, $P^1_{UV}$ inherits 
		the direction of monotonicity
		of $f$, while $P^0_{UV}$ has the
		opposite direction (such a direction for each coordinate exists as $f$ 
		is unate). 
		
		To deal with this, let $d \in \bbF_2^k$ be such that $u \mapsto f(u 
		\oplus d, w)$ is coordinate-wise non-decreasing for every $w$, and let 
		$\tilde U = U \oplus d$.  Then, $P^h_{\tilde U V}$ for $h = 0,1$ 
		exactly satisfy the monotonicity assumption of Proposition~\ref{prop: 
		gen_trunc} and hence, $\tilde U \better \tilde U \oplus V$.  Since 
		$\tilde U = U \oplus d$ is an invertible transformation of $U$ and 
		$\tilde U \oplus V = U \oplus V \oplus d$ is an invertible 
		transformation of $U \oplus V$, we also have $U \better U \oplus V$, 
		i.e., $Z_I \better AZ$, completing the proof.
	\end{proof}
	
	Thus, we have shown that to evaluate unate functions from binary linear 
	sketches, the best such sketch is simple truncation. 
	It is easy to see that this no longer holds if $f$ is not unate: consider 
	the simple example $f(z_1,z_2) = z_1 \oplus z_2$.  Then the binary linear 
	sketch obtained using $A = [1 \ \ 1]$ exactly recovers $f(z)$ for all $z$, 
	while truncation to either $z_1$ or $z_2$ cannot do better than an error 
	probability of $1/2$ for inputs uniform on $\bbF_2^2$.
	
	\section{Suboptimality of linear sketches for connectivity testing} 
	\label{sec: graph}
	
	We now use Theorem~\ref{thm: unate} to give a simple example where linear 
	sketches can be shown to be significantly worse than arbitrary functions, 
	by using truncation as an easy intermediary.  Linear sketches are commonly 
	used in distributed settings, and one such problem is that of testing graph 
	connectivity~\cite{ahn2012graph,yu2021tight}.
	We consider a one-way communication version of the problem:  Alice and Bob 
	observe graphs $G$ and $H$ respectively on the same vertex set, and are 
	unaware of each other's graphs (otherwise the problem is trivial).  Alice 
	sends a message as a function of $G$ to Bob, who wants to decide whether $G 
	\cup H$ is connected.  
	Sun and Woodruff~\cite{sun2015tight} showed that randomized
	one-way protocols for this problem require $\Omega(n \log n)$
	bits to have error probability (over the randomness of the protocol) at 
	most some small constant, say $1/10$, for every
	pair $(G,H)$ (and this lower bound can be achieved by considering a 
	deterministic protocol, namely sending a spanning forest of 
	$G$~\cite{hajnal1988graph}).
	
	In our example, we consider a (partial) distributional version where $G$ is 
	a uniformly random subgraph of a specific complete bipartite graph (as 
	described below), and we want the error probability (over the randomness of 
	$G$) to be at most $1/10$ for every $H$.  We show that truncation (and 
	hence all binary linear sketches) still requires $\Theta(n \log n)$ bits, 
	while we can construct a simple nonlinear message that does so with $O(\log 
	n)$ bits.
	
	More precisely, we consider a specific bipartite vertex set with $v$ left 
	vertices and $2^v$ right vertices, giving a total of $n = v + 2^v$ vertices 
	and $m = v 2^v$ possible edges.  Alice's graph $G$ has each of the $m$ 
	edges present independently with probability $1/2$.  Bob's graph $H$ can be 
	any fixed graph on the same vertex set\footnote{For completeness, it can be 
	checked that allowing both $G$ and $H$ to be uniformly random on this 
	vertex set makes the problem trivial: $G \cup H$ is then connected with 
	probability approaching 1, so always answering ``connected'' achieves 
	vanishing error probability with zero communication.}.  We represent $G$ 
	using its edge-indicator vector $Z \in \bbF_2^m$, with $Z_i = 1$ if the 
	$i$-th edge is present.  A binary linear sketch that is a $k$-bit message 
	is then $AZ$ for some rank-$k$ matrix $A$.
	Given a message $S$ and a decoder $g$ that maps $(S,H)$ to 1 for 
	``connected'' and 0 otherwise, define the error probability\vspace*{-5pt}
	\begin{equation}
		\grapherr{H}{S}{g} \coloneqq \Pr\left\{g(S,H) \neq \indic\{G \cup 
		H\text{ is connected}\}\right\}. \label{eqn: graph_err}
	\end{equation}
	
	\begin{theorem}[Graph connectivity testing]  \label{thm: graph}
		Let $G$ be a uniform subgraph of the complete bipartite graph with $v$ 
		left and $2^v$ right vertices, giving a total of $n = v + 2^v$ vertices 
		and $m = v2^v$ possible edges.  Then, for all sufficiently large $v$, 
		the following are true:
		\begin{itemize}
			\item[(i)]  There exists a deterministic protocol $\varphi$ using 
			$3v + 3 = O(\log n)$ bits and a decoder $g$ such that 
			$\grapherr{H}{\varphi(Z)}{g} \leq 1/10$ for every graph $H$ on the 
			same vertex set. 
			\item[(ii)]  On the other hand, if $\varphi(Z) = AZ$ is a $k$-bit 
			linear sketch of $Z$, then to have $\grapherr{H}{AZ}{g} \leq 1/10$ 
			for every $H$, it is necessary that $k \geq m/2 = \Omega(n \log n)$.
		\end{itemize}    
	\end{theorem}
	\begin{proof}
		\textbf{Upper bound in (i).}  We provide the claimed deterministic 
		protocol here and relegate its analysis to Appendix~\ref{app: 
		nonlin_ub}.  If there are at most three isolated right vertices and all 
		remaining (right and left) vertices form one connected component, then 
		Alice sends the labels of the isolated right vertices.  Otherwise, 
		Alice declares failure using a special message.  Upon receiving a 
		non-failure message from Alice, Bob immediately knows the component 
		partition of $G$ (he knows the locations of the isolated vertices and 
		that the remaining are connected), and is thus able to determine 
		whether $G \cup H$ is connected or not.  In Appendix~\ref{app: 
		nonlin_ub}, we show that the probability of Alice sending a failure 
		message is at most $1/10$ for sufficiently large $v$.
		
		\textbf{Lower bound in (ii).}  It is enough to show the lower bound for 
		a specific $H$; we take the empty graph $H = \varnothing$. Thus, Bob 
		wants to evaluate $f(Z) = \indic\{G \text{ is connected}\}$.  Since $f$ 
		is monotone and hence unate, Theorem~\ref{thm: unate} says that 
		truncation to an information set $I$ of $A$ performs at least as well 
		as $AZ$. Thus, for every decoder $g$, we have
		\begin{equation*}
			\grapherr{\varnothing}{AZ}{g} \geq \minerr{f}{AZ} \geq 
			\minerr{f}{Z_I},
		\end{equation*}
		where $\grapherr{\varnothing}{AZ}{g}$ is the error probability of using 
		decoder $g$ with the linear sketch to test if $G$ is connected (as 
		defined in \eqref{eqn: graph_err}), while  $\minerr{f}{AZ}$ is the 
		minimum error probability over all decoders to test if $G$ is connected 
		(as defined in \eqref{eqn: min_err}).  Hence, it suffices to obtain a 
		lower bound to the error probability of truncation.  To do so, let $p = 
		\Pr\{G \text{ is connected}\} = \bbE[f(Z)]$, and let the random 
		variable $Q = \bbE[f(Z) \mid Z_I]$.  Note that the smallest possible 
		error probability conditioned on $Z_I$ is $\min\{Q,1-Q\}$, and hence, 
		$\minerr{f}{Z_I} = \bbE[\min\{Q, 1-Q\}]$.  Simply lower bounding 
		$\min\{a,b\}$ for $0 \leq a,b \leq 1$ by $ab$, we have
		\begin{equation*}
			\minerr{f}{Z_I} \geq \bbE[Q(1-Q)] = \bbE[Q] - \bbE[Q^2].
		\end{equation*}
		The right-hand side, with some rearrangement and using the fact that 
		$\bbE[Q] = \bbE[f(Z)] = \bbE[f(Z)^2]$ as $f(z) \in \{0,1\}$, can be 
		written as $\Var(f(Z)) - \Var(\bbE[f(Z) \mid Z_I])$.  Hence, to 
		complete the proof, it suffices to show that the variance of $f(Z)$ 
		that is explained by $Z_I$, namely $\Var(\bbE[f(Z) \mid Z_I])$, cannot 
		be too close to $\Var(f(Z))$.  The desired bound is given by 
		Lemma~\ref{lem: var_red}, which is a specialized version 
		of~\cite[Theorem~1.1]{galicza2024sparse}.  We give a self-contained 
		proof for Lemma~\ref{lem: var_red} in Appendix~\ref{app: var_red}.
		
		\begin{lemma}[
			name={Variance reduction
				({\cite[Theorem~1.1]{galicza2024sparse}};
				proof in Appendix~\ref{app: var_red})},
			restate={[name={Variance reduction
					({\cite[Theorem~1.1]{galicza2024sparse}})}]varredlemma}
			]   \label{lem: var_red}
			Let $G$ be a uniformly random subgraph of a complete
			bipartite graph with $m$ possible edges, and let
			$Z \in \bbF_2^m$ be its edge-indicator vector.
			For $f(Z) = \indic\{G\text{ is connected}\}$ and
			every $I \subseteq [m]$,
			\begin{equation*}
				\Var\left(\bbE[f(Z) \mid Z_I]\right) \leq \frac{|I|}{m} 
				\Var\left(f(Z)\right).
			\end{equation*}
		\end{lemma}
		
		Note that $\Var(f(Z)) = \bbE[f(Z)] - \bbE[f(Z)]^2 = p - p^2$.  Thus, 
		using Lemma~\ref{lem: var_red}, we have that for $|I| = k$, 
		\begin{equation*}
			\grapherr{\varnothing}{AZ}{g} \geq \left(1-\frac{k}{m}\right)p(1-p).
		\end{equation*}
		We show in Appendix~\ref{app: graph_conn} that $p = \Pr\{G \text{ is 
		connected}\}$ goes to $1/e$ as $v \to \infty$, and hence, for 
		sufficiently large $v$, we have $p(1-p) > 1/5$ as $1/e(1-1/e) > 0.23 > 
		1/5$, so $k \geq m/2$ is necessary for the lower bound to be smaller 
		than $1/10$.
	\end{proof}
	
	Thus, we have an example where the limitations of linear sketches can be 
	easily established by deferring to the proxy of truncation: arbitrary 
	nonlinear functions can achieve a small constant error using only $O(\log 
	n)$ bits of communication, but truncation (and consequently, by 
	Theorem~\ref{thm: unate}, all linear sketches) requires $\Theta(n \log n)$ 
	bits, where $n$ is the number of vertices of the graph.

	\newpage
	
	\section*{Acknowledgements}
	A.G.\ would like to thank Amit Rajaraman for a careful reading of the 
	introduction and for providing feedback that improved clarity and 
	organization.
	
	\section*{AI methodology}
	Coming from an information theory background and being unfamiliar with the 
	broad theoretical computer science literature, the authors used AI tools 
	(OpenAI's GPT-6.1 ``Sol'') to help with identifying and making sense of 
	related work, particularly for the presentation in the introduction.  These 
	tools were also used to identify the applications to graph connectivity 
	testing and evaluation of Boolean functions (which were selected from a 
	long list of suggestions and details worked out by the authors).  All 
	theoretical results are due to the authors directly, as is all of the 
	writing.  GPT-6.1 was also used to identify typos and grammatical errors, 
	which were subsequently corrected by the authors.

	\printbibliography
	
	\newpage
	\appendix
	
	\section{Appendix: Technical proofs for graph connectivity analysis}
	
	Here we provide details of technical intermediate results that were skipped 
	in the main body of the paper to allow for a coherent presentation of the 
	most relevant ideas.
	
	\subsection{Analysis of nonlinear protocol for graph connectivity} 
	\label{app: nonlin_ub}
	We complete the proof of the upper bound in Theorem~\ref{thm: graph} by 
	showing that the nonlinear protocol described achieves an error probability 
	at most $1/10$ for sufficiently large $v$, using $3v + 3 = O(\log n)$ 
	bits.   The proposed deterministic protocol is as follows:  If there are at 
	most three isolated right vertices and all remaining (right and left) 
	vertices form one connected component, then Alice sends a leading $0$, 
	followed by two bits to specify the number of isolated right vertices 
	(which is one of $0,1,2,3$), and then the labels of the isolated right 
	vertices (this requires $v$ bits each; she sends at most three).  
	Otherwise, Alice declares failure by simply sending a single bit $1$.  Upon 
	receiving a non-failure message (i.e., not the single bit $1$) from Alice, 
	Bob immediately knows the component partition of $G$ (he knows the 
	locations of the isolated vertices and that the remaining are connected), 
	and is thus able to correctly determine whether $G \cup H$ is connected or 
	not.  Bob can declare arbitrarily upon receiving a failure message from 
	Alice.
	
	Thus, it is enough to bound the probability that Alice declares failure and 
	show that it is at most $1/10$ for sufficiently large $v$.   As the $v$ 
	edges from a given right vertex are chosen uniformly randomly, the 
	probability that a right vertex is isolated is $1/2^v$, independently of 
	the other right vertices.  Hence, the probability that any fixed four right 
	vertices are all isolated is $1/2^{4v}$.  Taking a union bound over all 
	choices of four right vertices, we have
	\begin{equation*}
		\Pr\{\text{there are at least }4 \text{ isolated right vertices}\} \leq 
		{2^v \choose 4}  \frac{1}{2^{4v}} = \frac{2^v(2^v-1)(2^v-2)(2^v-3)}{24 
		\cdot 2^{4v}},
	\end{equation*}
	which goes to $1/24$ as $v$ goes to infinity.
	
	Next, we consider the probability that the remaining vertices do not form a 
	single connected component.  Fix two distinct left vertices.  Each right 
	vertex is connected to both these left vertices with probability $1/4$, 
	hence the probability that this pair of left vertices has no common 
	neighbour is $(3/4)^{2^v}$.  Taking a union bound over all pairs of left 
	vertices, we have that 
	\begin{equation}
		\Pr\{\text{there is a pair of left vertices with no common neighbour}\} 
		\leq {v \choose 2} \left(\frac{3}{4}\right)^{2^v}, \label{eqn: 
		left_vertices}
	\end{equation}
	which also goes to zero as $v$ gets large.
	If all pairs of left vertices have a common neighbour, then all left and 
	non-isolated right vertices belong to the same connected component.  Hence, 
	the probability that Alice declares failure is at most the sum of the 
	above, which goes to $1/24$ and is thus smaller than $1/10$ for 
	sufficiently large $v$.   \hfill $\qed$

	\subsection{Proof of variance reduction lemma} \label{app: var_red}
	
	We restate Lemma~\ref{lem: var_red} then provide a self-contained proof.
	\varredlemma*
	\begin{proof}
		We use techniques from Fourier analysis, 
		see~\cite{ODonnell_2014,kannan2018linear} for a more detailed 
		exposition.  For a subset $J \subseteq [m]$, define the character 
		$\chi_J(z) = \prod_{\ell \in J}(-1)^{z_\ell}$.  Characters form an 
		orthonormal basis for the set of functions from $\bbF_2^m$ to $\bbR$, 
		in particular, $\bbE[\chi_J(Z) \chi_K(Z)] = \indic\{J = K\}$ for $J,K 
		\subseteq [m]$.  
		The Fourier expansion of $f$ is
		\begin{equation*}
			f(z) = \sum_{J \subseteq [m]} \hat f(J) \chi_J(z)
		\end{equation*}
		where $\hat f(J)$ are the Fourier coefficients of $f$, which are 
		real-valued.  Observe that $\bbE[\chi_J(Z) \mid Z_I]$ for $I \subseteq 
		[m]$ is equal to $\chi_J(Z)$ if $J \subseteq I$ and $0$ otherwise, as 
		the unconditioned uniform bit is averaged over.  Hence, by linearity of 
		expectation, we have
		\begin{equation*}
			\bbE[f(Z) \mid Z_I] = \sum_{J \subseteq I} \hat f(J) \chi_J(Z);
		\end{equation*}
		the only change from above being the restriction to $J \subseteq I$ (in 
		particular, if $I = [m]$, we have $\bbE[f(Z) \mid Z] = f(Z)$ and if $I 
		= \varnothing$ is the empty set, we have $\bbE[f(Z)] = \hat 
		f(\varnothing)$).
		We now compute the variance of the conditional expectation 
		$\Var(\bbE[f(Z) \mid Z_I])$, i.e., 
		\begin{equation}
			\Var\bigg(\sum_{J \subseteq I} \hat f(J) \chi_J(Z)\bigg) = \sum_{J 
			\subseteq I} \Var\big(\hat f(J) \chi_J(Z) \big) =  
			\sum_{\varnothing \neq J \subseteq I} \hat f(J)^2, \label{eqn: 
			var_cond-exp}
		\end{equation}
		thanks to the orthonormality of $\chi_J$ (with $\varnothing$ excluded 
		as $\chi_{\varnothing} = 1$ and has zero variance).  In particular, 
		taking $I = [m]$ gives $\Var(f(Z)) = \sum_{J \neq \varnothing} \hat 
		f(J)^2$.
		
		We now compare the two variances, for which it will be useful to 
		define, for edge $\ell \in [m]$, the quantity $a_\ell \coloneqq \sum_{J 
		\ni \ell} \hat f(J)^2/|J|$.   First note that the sum of all $a_\ell$'s 
		is $\Var(f(Z))$, as
		\begin{equation}
			\sum_{\ell \in [m]} a_\ell = \sum_{J \neq \varnothing} \sum_{\ell 
			\in J} \frac{\hat f(J)^2}{|J|} = \sum_{J \neq \varnothing} \hat 
			f(J)^2.   \label{eqn: var_expanded}
		\end{equation}
		Since the edges of a complete bipartite graph are symmetric, $a_\ell = 
		\Var(f(Z))/m$ for each $\ell$.  
		On the other hand, the variance of the conditional expectation is, by 
		\eqref{eqn: var_cond-exp},
		\begin{equation}
			\sum_{\varnothing \neq J \subseteq I} \hat f(J)^2 = 
			\sum_{\varnothing \neq J \subseteq I} \frac{\hat f(J)^2}{|J|} 
			\bigg(\sum_{\ell \in J} 1\bigg) =  \sum_{\ell \in I} \sum_{J 
			\subseteq I, J \ni \ell} \frac{\hat f(J)^2}{|J|}. \label{eqn: 
			var_cond-exp_expanded}
		\end{equation}
		Observe that the term being summed on the right-most expression in 
		\eqref{eqn: var_cond-exp_expanded}, for each $\ell \in I$, is at most 
		$a_\ell$, and hence, $\Var(\bbE[f(Z) \mid Z_I])$ is at most $\sum_{\ell 
		\in I} a_\ell = |I|/m \Var(f(Z))$.  This completes the proof.

		The above definition of $a_\ell$ can be motivated as follows.  
		For a nonempty set $J \subseteq [m]$, ``split'' its weight $\hat 
		f(J)^2$ equally among the $|J|$ edges in $J$.  For edge $\ell \in [m]$, 
		$a_\ell$ is the total weight assigned to edge $\ell$ from all sets $J$ 
		by such a splitting.  Summing $a_\ell$ over all edges $\ell \in [m]$ 
		yields the full variance $\Var(f(Z))$, hence $a_\ell$ can be considered 
		the contribution of edge $\ell$ to the total variance of $f$.
		However, for a given $I$, the variance of the conditional expectation 
		$\Var(\bbE[f(Z) \mid Z_I])$ includes contributions from edges $\ell \in 
		I$ that are \emph{not} $a_\ell$ (which includes shares of weights from 
		\emph{all} sets containing $\ell$), but rather, only the shares of 
		weights from sets containing $\ell$ that are also themselves contained 
		in $I$.  Thus, summing $a_\ell$ over all edges $\ell \in I$ gives an 
		upper bound to $\Var(\bbE[f(Z) \mid Z_I])$.
	\end{proof}
	
	\subsection{Probability of random subgraph connectivity} \label{app: 
	graph_conn}
	As stated in the proof of Theorem~\ref{thm: graph}, we show that the graph 
	$G$ (a uniformly random subgraph of the complete bipartite graph with $v$ 
	left vertices and $2^v$ right vertices) is connected with probability 
	approaching $1/e$ as $v \to \infty$.  
	
	We showed in Appendix~\ref{app: nonlin_ub} that the probability that all 
	left and non-isolated right vertices do NOT all belong to the same 
	connected component is at most ${v \choose 2}(3/4)^{2^v}$, which goes to 
	zero as $v \to \infty$ (see \eqref{eqn: left_vertices}).  Hence, in the 
	limit as $v \to \infty$, the probability that $G$ is connected is simply 
	the probability that $G$ has no isolated right vertices.
	Each of the $2^v$ right vertices is isolated with probability $1/2^v$ 
	(independently), as each of its $v$ edges is picked with probability 
	$1/2$.  Hence, the probability that $G$ has no isolated right vertices is 
	$(1-1/2^v)^{2^v}$, which has limit $1/e$ as $v \to \infty$, and we are done.

	\section{Appendix: Notation summary} \label{app: notation}
	For completeness, here is notation that we use throughout the paper.  We 
	write $[n] = \{1,\dots,n\}$ and $[t]_+ = \max\{t,0\}$. $\bbF_2$ denotes the 
	field given by the set $\{0,1\}$ with addition $\oplus$ and multiplication 
	modulo 2.
	Uppercase italicized letters play multiple roles: $S,U,V,W,X,Y,Z$ refer to 
	random variables; $A,B$ refer to matrices; $G,H$ refer to graphs; $I,J,K$ 
	denote coordinate index sets (which are subsets of $[n]$ or $[m]$); $K$ 
	refers to a transition kernel, represented by a stochastic matrix 
	(non-negative entries with rows summing to one). 
	Given a vector $x \in \bbF_2^n$, a matrix $A \in \bbF_2^{k \times n}$ and 
	subset $I \subseteq [n]$, $x_I$ denotes
	the subvector indexed by $I$ and $A_I$ denotes the submatrix formed by
	the columns indexed by $I$. $I^c$ denotes the complement of set $I$, which 
	in our case is always $[n] \setminus I$. The boldface letter $\eye$ denotes 
	the identity matrix (which is always $k \times k$ here), while $\indic$ 
	denotes the indicator function that evaluates to 0 or 1 accordingly as its 
	argument is false or true.
	
	$P,Q$ refer to probability distributions with subscripts possibly used to 
	clarify the random variable when important (e.g., $P_X$ is the probability 
	distribution of random variable $X$) and superscripts possibly denoting a 
	hypothesis $h = 0,1$.
	Calligraphic letters (e.g.~$\calW,\calZ$) denote the alphabet that a random 
	variable (e.g.~$W,Z$) takes values in (with the exception of $\calR_{k,n}$ 
	in Section~\ref{sec: future}).  Blackwell dominance is represented using 
	$\better$.  
	We use $\err{}$ to denote various notions of error, while $e$ denotes the 
	standard mathematical constant.  We write $\Df{f}(\cdot\,\|\,\cdot)$ and 
	$\Egamma{\gamma}(\cdot\,\|\,\cdot)$ to denote the $f$- and 
	$\Egamma{\gamma}$-divergences between two distributions.  We use 
	$\Pr\{\cdot\}$ to denote the probability of an event and $\bbE$ and $\Var$ 
	to denote the expectation and variance, all computed over the joint 
	distribution of all random variables involved.  We use $\Bern(p)$ to refer 
	to the Bernoulli distribution with bias $p$: $Z$ distributed as $\Bern(p)$ 
	(abbreviated as $Z \sim \Bern(p)$) takes value 1 with probability $p$ and 0 
	otherwise.  The product distribution is denoted as $\Bern(p)^{\otimes n}$: 
	$Z \sim \Bern(p)^{\otimes n}$ means that $Z = (Z_1,\dots,Z_n)$ with each 
	$Z_i$ independently and identically distributed (i.i.d.) as $\Bern(p)$.

\end{document}